# Democracy: Order out of Chaos
## Understanding Power-Law in Indonesian Elections

**Hokky Situngkir**
(hokky@elka.ee.itb.ac.id)
Dept. Computational Sociology
Bandung Fe Institute

**Yohanes Surya**
(yohaness@centrin.net.id)
Dept. Physics
Universitas Pelita Harapan

**Abstract**

We construct a majority cellular automata based model to explain the power-law signatures in Indonesian general election results. The understanding of second-order phase transitions between two different conditions inspires the model. The democracy is assumed as critical point between the two extreme socio-political situations of totalitarian and anarchistic social system – where democracy can fall into the twos. The model is in multi-party candidates system run for equilibrium or equilibria, and used to fit and analyze the three of democratic national elections in Indonesia, 1955, 1999, and 2004.

## 1. Our understanding on Democracy

What is democracy? If each citizen can vote the leaders or representatives according to her own opinion, can the result of the voting or election show the sense of democracy? For some democratic transitional countries, the questions are very important to have answers, since a 'democratic election' is believed to be the very first step to a democratic regime.

In recent work, Situngkir & Surya (2004) proposed an alternative way to extract information from the data of the general election in Indonesia and showed the clustering among political parties based upon the political stream realized as the fundamental for each political party. The political actions based on ideological streams in Indonesia have been accepted even since the first democratic election 1955 (Feith, 1970). The fact implies that the success of a political party depends upon the networks of the social organizations circling certain political parties.

The voters are not very free even in the democratic regime, since individual choices depend very much on the choices of the social networks where the voter embeds. However, this is natural for human being, since the different social identities may result social tensions, thus the micro-social is attracted to reduce the tensions by adjusting the political choices or ideological streams (Lustick & Miodownik, 2002).

The probability of a newborn political party to become majority is extremely hard but may occur in strangely special occasions. There have been some agent-based and Monte-Carlo models on how a minor political candidates can eventually gain a significant votes, e.g. Sznajd model (Stauffer, 2001) that simulating based on the Ising spin model – in the spatial model, certain number of agents persuade their neighbors to have the same political choices. In the other hand, some models inspired from the cellular automata showed some different rule patterns to understand social complex dynamics (Hegselmann & Flache, 1998). Moore (1996) shows some computational facts of such majority-vote 3-dimensional cellular automata dynamics.

The paper presents a little modification on majority vote cellular automata. The basic idea is to understand the microstructure of voters whose macro-properties showed facts on political streams turning out from the circling of societal identity. We construct the spatial model of virtual world in which agents choices depend much on the political streams of neighbors. Eventually, we showed that the numbers of neighbors become important variables presenting the macro-properties showing democracy as a critical situations among the extreme totalitarianism and anarchic society.

## 2. The Locality of the Voters Model to the Landscape of Voting

Individual voter is symbolized as a cell or square lattice located at a two-dimensional virtual world of $x, y = 1,...,n$ and she chooses any political party of $c \in C = \{c_1, c_2, ..., c_m\}$, where $m$ is the number of candidates. Aggregately, all voters choose $c_i$ can be stated as:

$$V(t)\big|_{c_i} = \sum_{x,y} v_{x,y}(t)\big|_{c_i} \qquad (1)$$

The change of votes (states) from one choice to another as a transformation from $t$ to $t+1$:

$$v_{x,y}(t+1) = f\left(v_{x,y}(t), v_j(t)\big|_{j=1}^{r}\right) \qquad (2)$$

$$v_{x,y}(t+1) = c_i \text{ iff } c_i = \max\left(\left[v_{x,y}(t) \oplus v_j(t)\big|_{j=1}^{r}\right]_1^m\right) \qquad (3)$$

or

$$v_{x,y}(t+1) = majority\left[v_{x,y(t)}, v_j(t)\big|_{j=1}^{r}\right] \qquad (4)$$

where $j$ is the neighbor index, and $r$ is the number of the neighbors. In other words, the changes of the states depend on the recent states of immediate neighbors and her own. For each neighbourhood of the adjacent cells, the next state of certain cell is decided by a majority vote among herself and the neighbours. The neighbours, conducting as the political circle around individuals force agent to change her state to the majority state. Technically, we use several modes of neighbourhood, as described in figure 1.

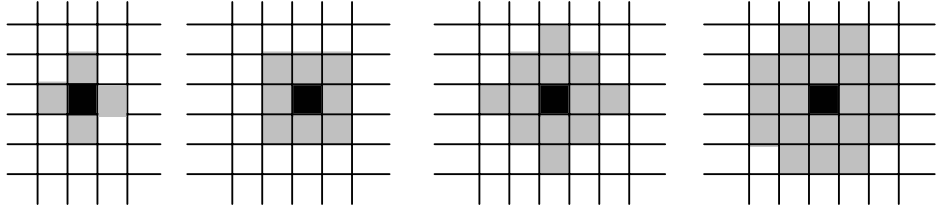

**Figure 1**
Various type of neighborhood (*left-right*): Von Neumann, Moore, extended Moore (4 added), extended Moore (12 added) neighborhood.

The more the number of the neighbors, the less the political freedom owned by an agent.

Consequently, the probability of changes to certain political parties, $c$, is

$$\Pr(c \to c_i) = \frac{\Delta V(t)}{V(t)}\bigg|_{c_i} \qquad (5)$$

where

$$\Delta V(t)\big|_{c_i} = V(t+1)\big|_{c_i} - V(t)\big|_{c_i} \qquad (6)$$

as the addition or subtraction of voters in each round of the game.

Eventually, the virtual world where the voters laid on consists of lattices and grids, while the global view is more like a torus (figure 2); the lowest two-dimensional grids are pasted together with the highest, and the right grids with the left one.

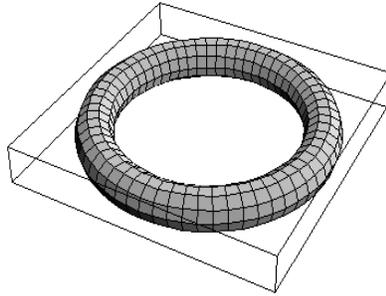

**Figure 2**
The torus as the model of the virtual political world

We do several simulations by using the rule explained above and discover that after several iterations the lattices are clustered altogether with neighborhood votes for the same choices (figure 3) – we denote it as an equilibrium condition (or equilibria[1]) that sensitive to the initial condition since the square lattices are not going to changes any further.

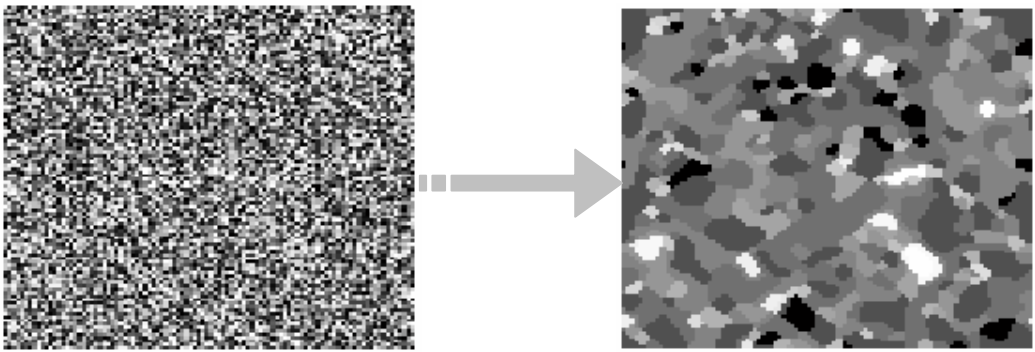

**Figure 3**
The initial condition (uniformly distributed votes among people) and the eventual one (clustered votes among people) for 20 competitive political parties.

Along the iterations, the square lattices organize themselves to the equilibrium. An example of our simulations is described in figure 4. The self-organized square lattices run for the equilibrium, states in which we can count as voting process; the voters organized themselves concerning their own choices and their neighbor's.

## 3. Democracy beyond totalitarian and anarchistic society

There are two extreme situations for each agent, i.e.: a totalitarian state in which every individual agent will turn to one totalistic state and in the other hand an anarchistic state in which individual does not take care about what her neighbors on their political choices – nobody count on the state. We can recognize the first extreme representation in a result of election of only one majority and the second one of uniformly distributed political choices that are no clustering among agents (figure 5).

---

[1] Several initial conditions do not stop at certain eventual condition but wavy or fluctuating tallies.

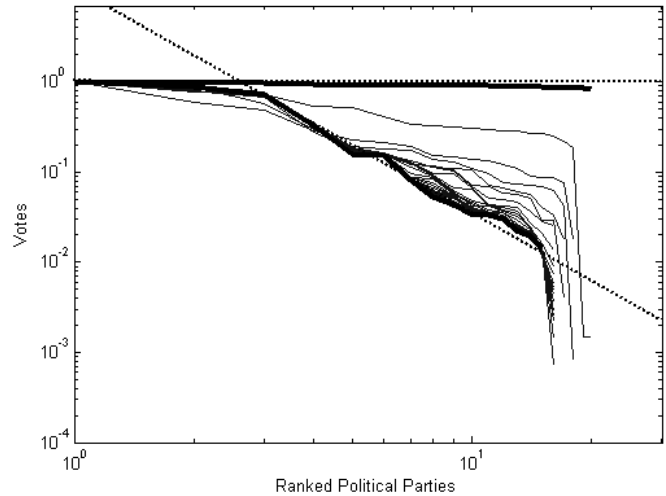

**Figure 4**
The ranked tallies of political parties in each round from *200* iterations over *100x100* square lattices of agents. It begins from the uniform distribution and eventually distribution presenting certain slope.

Thus, the born of a nation state can be understood as the way to dim the societal anarchy begun from the social contract among citizens (Rousseau, 1762; Hobbes, 1651). However, too much endeavor on damping the societal anarchy, the state may emerge the totalitarianism – a democratic state can turns out to be totalitarianism of majority rule (Tocqueville, 1840) or a dictator bureaucracies from the lack balances in governmental powers (Situngkir, 2003). Democracy is then laid in critical points between those two extreme situations. If the transitional phases from the anarchistic society to the totalitarian one are seen as second order phase transition, then we can say that the most critical system between the two extreme phases are the democracy – in a democratic election the citizens have the property of sustainable self-organized criticality. Second order transitions are transitions in a more gradual sense - on one side of the transition, a system is typically completely disordered, but when the transition is passed, the system does not immediately become completely ordered. Instead, its order increases gradually and evolves as the parameter is varied (Wolfram, 2002:981).

Since the transition from the two extreme phases represents the transition from a disorder level to the order one, then logically the critical points of democracy emerge the power-law properties (Schuster, 70-72). The understanding of the transitional states as an impact of self-organized in critical situations can be used in the analysis of statistical properties of election results in some countries like Indonesia to see how the nation state passes such transitions to assemble a more democratic system (Situngkir, 2004).

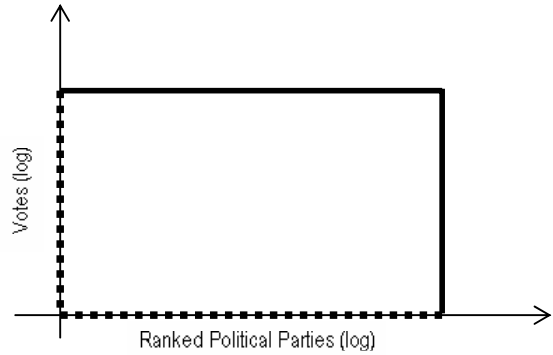

**Figure 5**
Recognizing the political situation (whether totalitarian - represented by dotted line - or anarchistic - represented by solid line) from the ranked political parties in the whole population.

Furthermore, in micro level the two extreme situations can be represented by using the parameter of the number of neighbors influencing a voter. In the first phase, the society is in the anarchistic social system, the average number of neighbors, denoted as *r*, can influence social agent is at the minimum ($r \to 0$) – agent acts and chooses the political state as she wants it to be privately and no social consensus exists. In the second phase, the totalitarianism rules in the society and there is only one majority since in the micro-level the agents are forced to have the uniform political streams by referencing the majority of the global ($r \to \infty$). Henceforth, the highest-level democracy lays in certain critical points of $0 \ll r \ll \infty$, of the power-law exponent around the unity, (*slope* $\neq \infty$ as vertical line and *slope* $\neq 0$ as horizontal line in the log-log plot).

**4. Simulations and Discussions**

As described in detail in Situngkir & Surya (2004), voters cluster in social institutions and organizations become the micro-property of nation-ship in Indonesia and shape the statistical properties of Indonesian election. This is the nature of Indonesian voters. The totalitarianism exists on the microstate depends on the global majority of the political system and the anarchistic society depends on the local majority of agents. Therefore, theoretically we can justify how statistically we have the mean of three democratic election ever had in Indonesia, 1955, 1999, and 2004. The normalized ranked votes on political parties in the three elections are different in the fitted slopes concerning the number of neighbors of each. The election 1955 as the most democratic election ever had in Indonesia is fitted with the von Neumann and Moore neighborhood while the 1999 with both of the extended Moore we use in simulation. The election of 2004 is fitted with the 2-agents neighborhood; an interesting result as compared to the recent political issues grew in the election 2004. All of the simulations conducted employ *100*x*100* square lattices run up to 200 iterations.

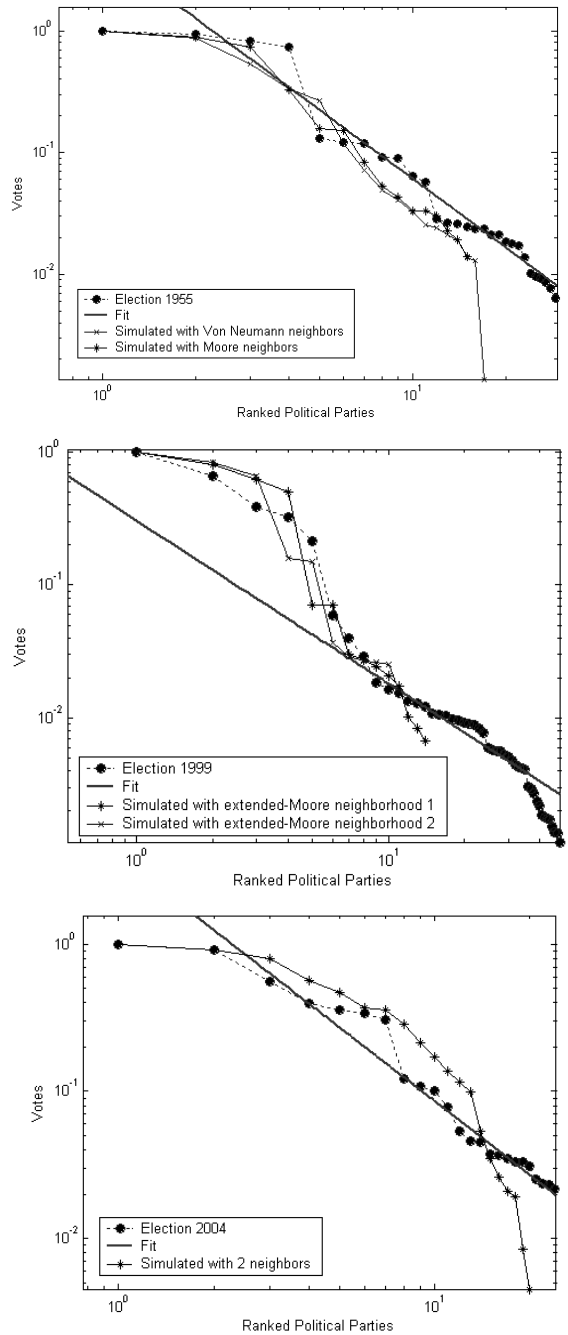

**Figure 6**
The normalized simulated data and the normalized real ranked tallies of political parties from the election 1955, 1999, and 2004 in log-log plot – different in number of neighbors.

However, the three election results reflect a self-organization among Indonesian voters to have democracy by holding general elections. The three figures plotted in the ranked political votes show us the evolution of the democracy in Indonesian nation-wide. Furthermore, we can see it from the figures how democracy grows based on different institutionalized ideological-streams and social identities.

**5. Concluding Remarks**

We present a spatial model captured the evolution of socio-political system evolves to democratic state. A fair and just election is an important milestone and important moment for every country to have national democracy. The spatial model presented incorporates the square lattices or cellular-automata based modeling. It is shown also that democracy can be assumed as critical points of self-organized agents in the transition from anarchistic society to the totalitarian one.

The model is implemented to the result of three democratic general elections held in Indonesia. The three national elections result the power-law signatures and fit with different types of neighborhood. Theoretically, the microstates of macro-properties transition of order to disorder and totalitarian to anarchistic can be approached as type of neighborhood chosen in the virtual world of social simulations.

The model can be useful to explain the power-law signature found in election results in Indonesia and furthermore to see how a democratic harmony evolves through heterogeneous social identities.

**Acknowledgement**

The authors thank Tiktik Dewi Sartika for discussions on Sznajd model and Surya Research Int'l. for funding the research. All faults remain the authors.